\documentclass{article}

\usepackage{PRIMEarxiv}
\usepackage{siunitx}
\usepackage[utf8]{inputenc} 
\usepackage[T1]{fontenc}    
\usepackage{hyperref}       
\usepackage{url}            
\usepackage{booktabs}       
\usepackage{amsfonts}       
\usepackage{nicefrac}       
\usepackage{microtype}      
\usepackage{lipsum}
\usepackage{fancyhdr}       
\usepackage{graphicx}       
\graphicspath{{media/}}     

\title{An Overview of Current and Emerging Biomaterials Technology for Continuous Glucose Monitoring (CGM) Devices - Current state and future perspectives of the leading technologies
\thanks{\textit{\underline{Citation}}: 
\textbf{Authors. Title. Pages.... DOI:000000/11111.}} 
}

\author{
  Umme Hafsa Momy \\
  Florida International University \\
  Miami\\
  \texttt{umomy001@fiu.edu}
}

\begin{document}
\maketitle

\begin{abstract}

In the world today, diabetic complications are a major factor in the disease's high mortality rate. Diabetes mellitus has been a major source of concern for decades due to its global prevalence and the resulting rising costs to individuals, governments, and healthcare systems on both a social and economic level. The complex interplay between nutrition, exercise, stress, medicine, sickness, and hormonal changes makes controlling diabetes difficult. Even though there is currently no cure for diabetes, the condition may be managed with careful attention to blood glucose levels and the use of appropriate medications. Blood sugar levels should be monitored on a regular basis, which requires a glucometer and a fingertip and is becoming the standard in diabetes control technologies. For the convenience of diabetic patients, the use of a CGM device for monitoring glucose levels has become the preferred method. For diabetes therapy at the point of care, replacing finger-prick glucometers can be done anywhere they are currently on the market. As a result, it has sparked a lot of interest in a variety of glucose sensing mechanisms, particularly non-invasive or marginally glucose sensing mechanisms that focus on subcutaneously implantable electrochemical sugar sensors that work more reliably and last longer.
\end{abstract}

\keywords{Continuous glucose monitoring; diabetes glucose biosensor; implanted devices; Blood glucose monitoring; artificial pancreas; mini-invasive; green initiative; non-invasive)}

\section{Introduction}

A worldwide epidemic known as diabetes mellitus causes chronic hyperglycemia as a result of either an inadequate supply of or a tolerance to insulin \cite{chen2022functional}. As a result, In the United States, diabetes ranks seven among main causes of mortality, responsibility for 270K deaths in 2017 and affecting 422 million people globally \cite{world2018global}. According to the CDC 2020 Report, 120 million Americans over the age of 18 currently have diabetes, and the prevalence of the disease is rising among children under the age of 18 as well. If current trends continue, that percentage will reach one in three adults by 2050 \cite{newman2005home}. Understanding the epidemic is improved by classifying the different types of diabetes. According to the CDC's 2020 report, of the world's 120 million people with diabetes, 32.5 million have type II, 1.71 million have type I, and the rest 88 million have prediabetes, commonly known as impaired glucose tolerance (IGT) \cite{cdc2023}. An annual cost burden of \$174 billion in the United States, However, \$327 billion was spent on diabetes in 2017, with \$237 billion going toward medical expenses and \$90 billion to lost productivity \cite{centers2011national}.

Diabetes patients' blood glucose (BG) levels can vary drastically all day, this might cause life-threatening side effects including as cardiac arrest, stroke, hypertension, renal failure, vision loss, coma, and amputation \cite{centers2011national,wild2004global,chase2010continuous}.

The first test strip and reader-based personal glucose testing system were introduced in 1987 by Medisense, Inc. Since then, more than 40 different commercials and portable monitors have been released \cite{newman2005home}. The mainstay of conventional Measurements of Glucose methods by using electrochemical techniques amperometric glucose sensors \cite{clark1962electrode}. More than 25 glucose monitors have received FDA (Food and Drug Administration) approval so far and most of them use test strips with they typically use

 glucose oxidase (GOx) or glucose dehydrogenase (GDH) immobilized on test strips to determine blood sugar levels on a screen-printed electrode \cite{wang2008electrochemical}. With the help of glucose sensors, people with diabetes can now check their own blood sugar levels to control their insulin levels and lower their risk of getting diabetes mellitus. When insulin is not secreted (type I diabetes, T1D) or is secreted but is not acting properly (type II diabetes, T2D), the BG concentration is abnormal. Patients with type 2 diabetes and type 1 diabetes, in that order, should have their blood glucose levels checked at least twice and four times day, respectively, along with a combination of therapies that includes medication, nutrition, exercise, and insulin injections all have a role in\cite{chase2010continuous,vigersky2015benefits,ehrhardt2011effect}.

Real-time glucose monitoring provided by CGM systems can assist physicians and diabetic patients in making well-informed healthcare decisions. Several challenges have been met with a he creation of glucose sensors appropriate for CGM use. Thus, it has been sought after to develop analytical techniques that permit continuous blood glucose monitoring \cite{malone2021combination}. In addition, although their clinical applicability is still debatable, the creation of monitoring sensors that are either less intrusive or not invasive at all that the presence of glucose in fluids such as weeping, perspiration, saliva, and urination has recently garnered a lot of attention. Continuous CGM offers real-time data on patterns, the size, the time and frequency of glucose changes throughout the day, i.e., whether the glucose levels are rising or falling \cite{malone2021combination,sun2021improving}. Even though these monitors have made it easier for persons with diabetes to regulate their sugar levels, they can only show the blood sugar level at the moment and cannot predict hyperglycemic or hypoglycemic events.

The majority of CGM systems can upload data to cloud databases or connect to a wide range electronic gadgets, including smartphones. The most notable feature of the CGM system is the alarming systems for hyper- and hypoglycemic levels. In other words, the current CGM system can predict how much glucose will be in the blood in the future. It can also keep track of glucose in real time in the body.

The best way to manage diabetes mellitus is through strict glycemic control because doing so lowers the risk of microvascular complications. The alarms for hyperglycemia and other conditions are the most prominent aspect of the CGM system. According to the research of the National Institute of Diabetes and Digestive and Kidney Diseases, which is part of the National Institute of Health (NIH-NIDKK).

\section{Evolution and foundational ideas of CGM systems}

\subsection{Clark and Lyons}
The Clark oxygen electrode was created by the late Professor Leland C. Clark Jr. of the University of Alabama Medical College, who also put forth the idea for the basic enzyme sensors. "An example of how this setup can be used is the measurement of glucose using a pH electrode and a Cuprophane-glucose oxidase Cuprophane membrane," Clark and Lyons wrote at the start of their first academic article, which explained two ideas.

Two fundamental ideas were introduced for the first time in the academic paper cited by Clark and Lyons in a joint effort, as follows: "The determination of glucose, using a Cuprophane-glucose oxidase Cuprophane membrane and a pH electrode, illustrates one use of this system." Gluconic acid is created when glucose diffuses through such a membrane and then diffuses back into the donor solution and toward the pH-sensitive glass. A system that is sensitive to glucose can be set up using a hydrophobic (e.g., polyethylene) dialysis membrane, glucose oxidase, a PO2 electrode, and a plasma membrane because oxygen is taken up from the moving glucose solution based on how much glucose there is in the solution. This causes the pH to drop, and how much it drops depends a lot on how much glucose is there.

The principle of potentiometric glucose enzyme activity is discussed in the earlier description. In the second case, the amperometric glucose enzyme sensor principle comes into play, which was described in his 1965 patent and is now called the first-generation glucose enzyme sensor.

\subsection{Updike and Hicks}
Then, Updike and Hicks released their first article in 1967 about a glucose sensor that used oxygen electrodes and glucose oxidase (GOx). These mark the beginning of the development of glucose and biosensors. Since then, the combination of biological elements and transducers has been reported in thousands of biosensor principles. Based on his original theory, the first electrochemical glucose sensor was put into production, but it used the hydrogen peroxide that was released when oxygen was reduced as the electron acceptor.

\subsection{Guilbault and Lubrano}
Guilbault and Lubrano established in 1973 that hydrogen peroxide generated enzymatically by glucose oxidase may be detected amperometrically \cite{vigersky2015benefits}. Yellow Spring Instruments introduced the first specialized glucose analyzer for direct detection of glucose in whole blood samples in 1975 \cite{clark1962electrode}. The most effective glucose concentration measurement methods to date have been based on enzyme-modified electrodes and electrochemical detection, with numerous useful designs.

Updike and Hicks created a sensor that detects hyperglycemia by measuring oxygen consumption using two oxygen electrodes (one covered with the enzyme and one for reference). The differential currents between these electrodes are compared to account for differences in oxygen background \cite{wang2008electrochemical}. The electrode design, electrode material, enzyme immobilization technique, and polymeric membrane compositions of amperometric enzyme electrodes vary widely. In 1982, Shichiri et al. created and tested the first continuous in vivo monitoring system for blood glucose. In 1999, the FDA gave its first premarket approval to a CGM system \cite{ehrhardt2011effect,malone2021combination}.


\section{Enzyme-immobilized electrodes serve as electrochemical glucose sensors:}

Electrochemical sensors based on enzymes have been the focus of most of the previous research on glucose sensors \cite{pickup2005fluorescence}. The idea of fixing GOx the electrode, which was initially suggested in the 1960s, has been the basis of research into this kind of biosensor \cite{bard2022electrochemical}.

Earlier glucose monitoring devices, which used the reflectometric principle, needed a substantial blood sample (about 30 l) to identify variations in the color of glucose test strips. Due to the presence of GOx, O2, and H2O in the first enzymatic glucose sensors, glucose was oxidized into gluconic acid and hydrogen peroxide. At the electrode, the glucose and catalytic enzyme might quickly react with oxygen, allowing for the purpose of measuring blood glucose levels by the detection of consumed O2 or the byproduct of oxidation \cite{clark1970membrane}. As adequate, there are concerns that the sensor would be significantly influenced by the environment of the blood sample due to the need for oxygen as the mediator and the inevitable interference of the electroactive species \cite{chase2010continuous}.

Glucose monitoring systems of the second generation, on the other hand, relied heavily on the electrochemical concept and included disposable electrode sensor tapes to allow for equipment downsizing with a negligible amount of blood sample. Using electrodes for enzymes that are disposable and portable sensors based on electrochemical reactions, which only need less than \SI{1}{\micro\liter} of whole blood, we were able to get results in a matter of seconds. The notion of an artificial mediator to supply an electron acceptor for the replenishment of oxygen was established in the second generation of enzymatic glucose sensors, and in the third generation, electrons are sent directly, as illustrated in figure 1 below. But there are still big problems, like a thick protein layer that makes it hard to move electrons and the fact that pH, temperature, and humidity in the human body can make enzymes work less well \cite{PICKUP20052555}.

 Amperometric detection is used in both the first and second generations of principle-based glucose sensors. In the first generation, hydrogen peroxide or oxygen reduction is monitored electrochemically, whereas in the second generation, reduced mediators are oxidized. Clark and Lyons (1962) stated that pH-sensitive electrodes or field-effect transistors (FETs) have been utilized in several investigations of potentiometric enzyme glucose sensors, which detect the generation of gluconic acid by measuring the resulting drop in pH. The amperometric glucose sensing technique was chosen by industries for SMBG sensors because it is inexpensive and can be used with disposable electrodes \cite{brockway2015fully,clark1959electrochemical}.


Using the 2nd generation sensor approach, a number of different GDHs were used. These included the nicotinamide adenine dinucleotide (NAD) dependent GDH, the pyrroloquinoline quinone (PQQ) dependent GDH, and the flavin adenine dinucleotide (FAD) dependent GDH \cite{ferri2011review}. The research published in 2011 \cite{ferri2011review}. Considering their oxygen insensitivity and higher substrate specificity, FADGDHs produced from fungus are now the most common enzymes for SMBG sensors \cite{OKUDASHIMAZAKI2020107414}. It has been shown \cite{OKUDASHIMAZAKI2020107414}. Because of its consistent stability and substrate specificity, GOx has remained the enzyme of choice in glucose sensors ever since the first glucose sensor was developed.

\section{Approaches of Keeping Control on Glucose}
\subsection{Evaluation of Individual Blood Samples}
Due to the widespread acceptance of blood glucose levels as the main diagnostic criteria for diabetes, several technologies were developed and commercialized for quick monitoring of glucose levels. The frequent evaluation of BG levels necessitates a small amount of blood (<1 \textmu L) that must be collected using the "finger-pricking" collection technique and then capillary action is used to inject a test substance into the test strip. This procedure is uncomfortable for the patient and has a low compliance rate \cite{newman2005home,wang2008electrochemical}. Furthermore, since instantaneous monitoring sensors are unable to provide real-time BG data, they cannot forewarn users in advance of hypoglycemic (low blood sugar, <3.0 mM) and hyperglycemic (high blood sugar >11.1 mM) events. Chemically, you can find out how much glucose is in your blood by pricking your fingertip and putting a little bit of blood on a disposable test strip \cite{clark1962electrode}.


However, similar examinations have been chosen to be conducted numerous times a day with discrete data collection, which is extremely uncomfortable and painful due to the inescapable frequency of intrusive testing that needs to be carried out routinely even when sleeping, driving, or in meetings \cite{clark1970membrane}. In addition, a diabetic's blood glucose level might fluctuate widely over the course of a short period of time, necessitating accurate real-time monitoring in order to accomplish optimal insulin administration. Glucose monitoring in real time requires taking blood samples more often than portable finger-prick blood glucometers can do.
\subsection{From stand-alone CGM to linking with insulin pump to realize artificial pancreas}

Initial versions of CGM procedures were only intended for professional use and when glucose meters stood on their own systems with no instantaneous capabilities. Healthcare professionals have been using professional CGM systems to measure and monitor diabetic patients' glycemic levels throughout their typical daily lives. The Menarini method is based on microdialysis and works by pumping a buffer solution via a catheter with a biocompatible membrane; the concentration of glucose in the effluent from the catheter may be monitored using an external monitor \cite{malone2021combination}. Improvements in the accuracy of glucose sensors and the creation of CGM system calibration techniques that use instruments for measuring glucose levels in the blood made it possible to use insulin pumps and CGMs together for insulin infusions under the skin at constant rates (CSII) therapy.

The development of sensor-enhanced insulin pumps, which are solely controlled by CGM system data, was the most notable accomplishment. The newest technology, the CGM system, which is part of the "Hybrid Closed Loop" system, is making progress toward achieving the artificial pancreas system. and the threshold suspend device system suspends the pump automatically before low glucose is anticipated. The greatest drawback is that the needed high flow rate to reduce lag time would result in inadequate time to achieve glucose equilibrium between the pumped fluid and the interstitial fluid (ISF) \cite{malone2021combination}. Also, because of the container with the medication and the pump, the system has to be bigger, which could be uncomfortable and make the patient feel less satisfied \cite{sun2021improving}.

\subsection{Electrochemical Glucose Implant Sensors for Continuous and Intermittent Monitoring}

Electrodes used in conjunction with enzymes that facilitate oxidation-reduction reactions (redox) may be able to detect a current or voltage that varies with enzyme concentration. Subcutaneous amperometric electrodes, microneedles, microdialysis, micropores, and intravenous implanted devices are all examples of electrodes used often in invasive sensor technologies. Just two types of devices, microdialysis and subcutaneous, have made it to market. Microdialysis devices employ a hollow microdialysis fiber to perfuse an ex vivo reservoir with isotonic fluid, while it is possible for IG to diffuse easily into the fiber and be pumped to an enzyme-based electrochemical sensor. It is easier to implant because of the system's compact size, and biofouling is less of an issue since the sensor is external to the body \cite{lee2021continuous}. When designing electrodes for subcutaneous procedures, the main goals are to make them more sensitive, selective, and biocompatible.

For subcutaneous implantation, the traditional needle-type sensor features a platinum-iridium (Pt-Ir = 0.125 mm, Pt:Ir = 9:1) wire working electrode with an immobilized mediator and enzyme, was a major advancement in biosensors for glucose in vivo in the past. An Ag/AgCl wire acts as a counter electrode and is wound around the working electrode. These gadgets are sensor implants meant to be put in humans, where they would function constantly for a few days before needing to be removed and replaced. Correction algorithms for the temporary dissimilarities (lag-time) between significant differences between BG and IG were studied for potential use. Union City, California-based firm Kumetrix has suggested a portable, battery-operated electronic monitor using silicon micro-needles on a scale comparable to that of human hair. The objective is to create a metering system compatible with the cartridge holding single-use sample tools. By doing so, a small quantity of blood will be obtained via the micro-skin needle's puncture (less than 100 nL).

Based on the creation of needle-type oxidase biosensors, coil-type oxidase biosensors were made. A 30-gauge needle was wrapped with a coil-shaped cylinder made of Pt-Ir wire measuring approximately 1 mm in length, 0.55 mm in width, and 0.3 mm in thickness. This layout significantly expanded the usable sensing surface of implanted biosensors and produced a three-inner-chamber thickness of 0.07 millimeters for more enzyme immobilization, all while staying within a practical length range. The sensor's consistent response could be maintained for at least 60 days with several tweaks, such as etching the coil's surface, applying hydrogel coatings, and loading it with particular medications \cite{brockway2015fully}.

\subsection{Implantable glucose monitors}

The application for an implanted CGM system has drawn attention to implantable technology shows a closed-loop system. It includes an implantable biosensor for glucose monitoring, a dosage control system, and a medication pump \cite{brockway2015fully}. Smaller devices with longer lifespans are often easier to implant. Foreign-body reaction, deterioration, and mechanical failure restrict the lifespan of implanted devices \cite{brockway2015fully}.

Because of the warm and salty environment, in vivo conditions may hasten the deterioration of implanted biosensors. Depicts how foreign-body reactions can disrupt sensor operation \cite{clark1962electrode}. The immune system's quest to eliminate the internal parasite occurs a few times after implantation owing to inflammation that occurs suddenly within the body. Biological substances, notably enzyme-based sensing compounds, degrade, generating a fibrotic capsule surrounding the implant. Within a month, metal and electrical gadgets would corrode and malfunction owing to water absorption, limiting their lifespan to 1–2 weeks \cite{brockway2015fully}.

To reduce the deleterious effects of implanted sensors, biocompatible glucose-sensitive biosensors have been developed. Hydrogels are popular because their mechanical and optical characteristics may be tuned \cite{wild2004global}. A boronic acid hydrogel membrane was attached to a quartz crystal using surface-initiated polymerization and its biocompatibility was microscopically validated. Within 7 days of implantation, no trace of effects on cells (cytotoxicity) and on infiltrating immune cells (inflammatory response) remained, and a 100-s reaction time could cover the whole range of blood glucose levels within a healthy range \cite{clark1962electrode}.

Boronic acid-functionalized hydrogel optical fiber has great performance and biocompatibility. Yetisen et~al. studied biocompatible hydrogel optical fibers that may include analyte-sensitive chemicals. With a change in glucose content, the hydrogel matrix swells and shrinks, causing variations in optical characteristics that may be readily identified \cite{yetisen2014holographic}. Glucose monitoring without the use of labels has enormous promise for implantable applications. Elsherif et al \cite{lee2021continuous} constructed a smartphone-integrated optical fiber for in vivo glucose sensing by photopolymerization, linking an asymmetric microlens-array imprinted glucose-sensitive hydrogel to a multimode silica fiber tip. This optical fiber is biocompatible; thus, it may be used to minimize inflammation at the implanted site and has a great compassion of 2.6 W m1 with a 15-minute reaction time and 0.1\% lactate interference \cite{lee2021continuous}.

Current implanted glucose sensors use glucose sensing via enzyme-based electrochemistry technologies. Even though CGM was accomplished using implanted sensors, they still face obstacles, including foreign-body reactions. Materials science has been used to minimize foreign-body reactions. When creating a new glucose sensor, biocompatibility was a priority. Hydrogel matrices are biocompatible. Thus, hydrogel optical fibers modified with boronic acid were produced and evaluated, exhibiting excellent sensitivity and biocompatibility for future implantable applications. In vivo testing is needed in search of further evidence.

\subsection{Non-invasive glucose testing}

Even though CGM systems provide continuous, instantaneous glucose readings with an insulin pump feedback system for effective checking of glucose levels, they have minimal patient compliance owing to their short lifespan and calibration requirements \cite{wild2004global}. Recently, glucose monitoring without needles or blood draws with wearable gadgets has gained popularity. Electrochemical or colorimetric sensors are used in noninvasive monitoring. Colorimetric sensors are visible to the human eye, whereas electrochemical sensors might readily offer continuous monitoring by delivering real-time data to wireless electronics \cite{clark1970membrane}.

The glucose concentration in ISF, rather than the actual measurement of glucose in the blood, is measured using Gluco Watch, which was the first non-invasive glucose monitor to hit the market \cite{Krans2018}. It is worn close to the wrist and uses an electrochemical sensor based on enzymes. Reverse iontophoresis, in which a negligible current is given to the skin to induce ion transfer, lies at the heart of this method \cite{wang2015recent}. At neutral pH, sodium ions are the primary transporters of charge; therefore, their movement would cause the ISF to flow electro-osmotically from the epidermis to the anode. In this way, glucose is carried to the cathode \cite{wang2015recent}. Standard glucose sensors, located at the cathode, determine glucose levels by direct enzymatic measurement (glucose oxidation by an enzyme, such as GOx). Using a glucose sample is one of the most studied ways to keep track of glucose because it is so accurate.

However, GlucoWatch's real-time glucose monitoring features are limited by its inability to correctly identify glycemic fluctuations in real time. One of the most significant unintended consequences is patient dissatisfaction, and one of the main reasons for this is skin irritation from the administered current \cite{Krans2018}. As a result, glucose monitoring through tattoos is an idea that has been utilized to circumvent this constraint by decreasing the amount of current needed used a temporary tattoo and an altered glucose oxidase with a lower threshold for Speakers that emit a bluish signal are called "Prussian Blue transducers" to measure electrolytes and metabolites in sweat(Bandodkar et al.) \cite{wang2015recent}. This makes it possible to have a platform that is flexible, cheap, and compatible with the body for continuous glucose monitoring without requiring a needle in the ISF \cite{Krans2018}.

The detection of glucose in other biofluids than ISF has also received considerable attention. Because it may be collected externally, sweat is an appealing biofluid for non-invasive applications because it would make this ongoing procedure more easy and pleasant for patients \cite{Krans2018}. From this concept, a number of glucose monitoring patches for continuous use have been developed. A thin, flexible patch that maintains skin contact and performs well even when physically deformed.  This patch incorporates gold-doped graphene, which enhances the electrochemical activity and hence the sensitivity of glucose monitoring. Glucose sensors have been linked to pH and humidity sensors for continuous calibration, which makes it less likely that the environment will affect the readings \cite{Krans2018}.

Many studies have focused on contact lens- and eyeglass-based detectors because, like sweat, tears are collectable biofluids externally. The contact lens-based glucose monitor reported by (Yao et al.) contains built-in electrochemical glucose sensors based on enzymes and a communication circuit, making it both wireless and capable of continuous monitoring (Yao et al. ) \cite{vashist2012non}.The wireless test setup used to describe the data Transmission Without Wires findings from glucose monitoring with a built-in sensor in a pair of contact lenses. Using a polydimethylsiloxane eye model, we described and evaluated the sensor, finding that it performed very similarly to tear fluid, indicating that it has significant promise for future use in continuous glucose monitoring without needles \cite{vashist2012non}. The restricted operating range for in vivo testing, as well as the polymer substrates' gas permeability for patient comfort all need more research \cite{vashist2012non}. Protein fouling and stability degradation throughout a period of time and at different temperatures are other areas that need more attention.

A non-invasive tear-sensing device has also been created lately that can detect various analytes, including glucose, from tears collected by a wearer's spectacles. In this case, a fluidic device is attached directly to the nose-bridge pad of the spectacles, allowing for direct collection of artificial tears \cite{vaddiraju2010technologies}. This eyewear framework's electrochemical sensing system is intended to be placed externally, With the external tiny flow detector's wireless electrical backbone incorporated into the frame, we may gather stimulated tears \cite{vaddiraju2010technologies}. This is done to avoid some of the problems with sensors that are built into contact lenses, such as low user compliance and the possibility that the embedded sensor system could make it hard to see.

Breath condensate and volatile organic molecules from exhaled air are two potential bio-fluids for non-invasive, continuous glucose monitoring \cite{moschou2004fluorescence}. The concentration of glucose in a healthy person's lung fluid is said to be only around 0.4 mmol l-1, which is about 10 times lower than the amount of glucose in the blood. A glucose monitor with minimal lag time may be possible because of the steep concentration gradient that forms between the respiratory fluid and the plasma. Due to the low glucose levels in respiratory fluid, the development of glucose monitors based on exhaled breath faces a significant roadblock \cite{moschou2004fluorescence}. Nonetheless, the sensibility and clarity of the presently existing glucose sensors should be enhanced.

The most investigated technique for monitoring blood sugar without causing pain is electrochemical observation of the epidermis, which is commercially accessible. This kind of glucose monitor measures biofluids such as skin ISF or perspiration instead of blood glucose. Glucose diffuses through the endothelium or sweat glands in blood arteries; the glucose concentration of these biofluids and blood glucose levels are related \cite{Krans2018}.

\section{Glucose Detection Based on Optical Approaches or Combination Technologies}

Infection risks, discomfort, and analytical response problems related to the FBR are characteristics of almost all invasive sensing devices \cite{MayoClinic2017,Krans2018}. Thus, research has also concentrated on the creation of minimally invasive and noninvasive spectroscopic methods to evaluate physiology that is accessible from the outside (such as skin, saliva, and tears) \cite{Krans2018}.

Fluorophores and non-fluorophores are two categories of glucose detection methods based on optical methods \cite{wang2015recent}. Fluorophore-based methods use a spectroscopic affinity sensor in which glucose and a fluorophore-labeled molecule compete for binding to a concanavalin A receptor, which is unique to both ligands \cite{do2008current,vashist2012non}.As an alternative, only glucose inhibits photoinduced electron transfer (PET) when it binds to a recognition site (such as boronic acid derivatives or a composite of fluorescent moieties on a hydrogel)\cite{vaddiraju2010technologies,moschou2004fluorescence}.

The concentration of glucose affects the fluorescence emission strength and/or lifetime of Förster resonance energy transfer (FRET) \cite{MayoClinic2017}. In this way, Heo et al. monitored interstitial glucose concentrations in a rodent model for 140 days using fluorescent hydrogel fibers (1000 m dia.) \cite{moschou2004fluorescence}. The hydrogel fibers, which were simple to remove, allowed transdermal optical detection of the fluorescence intensity \cite{pickup2005fluorescence}. Skin pigmentation and epidermal thickness can have a negative effect on the analytical response among hosts, even though optical detection offers sensitive detection without harming the host \cite{moschou2004fluorescence,pickup2005fluorescence}. Significant drawbacks include scattering in tissues and photobleaching of the fluorophore \cite{wang2015recent,moschou2004fluorescence,pickup2005fluorescence}. Personalized devices for continuous monitoring are also not currently possible due to the instrumentation's inability to be miniaturized \cite{wang2015recent}.

\subsection{Infrared spectroscopy}
Optical absorption spectroscopy is another potential glucose sensor. Spectroscopic techniques detect objects by measuring how much light they absorb, transmit, or emit \cite{unmussig2018non}. In glucose sensing, because biological tissues are too thick for transmission measurements, reflectance is examined instead. When light is directed towards the tissue's surface, a portion of it goes through the skin and re-emerges at the spot \cite{lim2005glucose}.

For implantable applications, most research has focused on MIR and NIR. An implanted glucose sensor that employs fiber optics and mid-infrared attenuated total reflection, as described by Li et al.\cite{lim2005glucose}.	It was built with a U-shaped construction to provide an expanded optical length of action inside the constrained space for implanted purposes and boost its glucose sensitivity. The sensor's sensitivity and resolution were increased by covalently adding silver nanoparticles \cite{lim2005glucose}. A NIR sensor on a chip utilized in CGM systems by Mohammadi et al. In vitro findings for glucose concentrations from 0– 400 mg dl1 were followed by in vivo measurements utilizing the NIR-CGM system, which demonstrated a detection limit of 20 mg dl1 and an average MARD of 13.8\% \cite{tu2003nanoelectrode}.

Absorption banding in MIR is caused by molecules flexing and stretching. Background adsorption bands create substantial interference \cite{kang2009glucose}. Overtone vibrations create NIR absorption bands. With a defined glucose peak, low water absorption bands may be obtained \cite{kang2009glucose}. 90p–95p of NIR light penetrates human skin, showing that this method accomplishes its goals with little tissue damage \cite{clark1970membrane}.

Similarities in the spectra of glucose and other sugar components are striking in tissue tension using this method, and tissue dispersing hampers measurement \cite{unmussig2018non}. The spectrum interacts with diffusion in the water's backdrop and skin location variant \cite{lim2005glucose}. To get around these problems, different types of spectroscopies were looked at, in particular Raman spectroscopy, which would provide sharp absorption peaks without relying on water.

\subsection{Raman spectrum}

Raman spectroscopy is one technique for detecting glucose. Raman-scattered light may be collected and detection system-filtered, delivering signs to the computer with a Raman shift. Raman spectroscopy may yield distinct bands according to concentration, although the overtone bands are audibly less prominent \cite{kang2009glucose}. Compared to NIR, Raman makes it simpler to detect target peaks, but the spectrum takes longer  \cite{clark1970membrane}. Due to its increased Raman signal responsiveness and superior sensitivity, surface-enhanced Raman scattering (SERS) is the most commonly used method. However, glucose's low surface adsorption capacity is still a problem \cite{kang2009glucose}.

The different studies that were done to improve glucose sensitivity and selectivity by using modified Raman-active substrates constructed a glucose-sensing device by electrostatically assembling GOx on a silver nanoparticle-functionalized SERS substrate \cite{kang2009glucose}. The whole range of human blood glucose, from 36 to 252 mg/dl1, was detected, which is a remarkable feat of sensitivity. After embedding bisboronic acid glucose sensors onto a gold-film-modified nanosphere substrate, found that the device showed promise for in vivo glucose monitoring with a detection range of 18–180 mg/dL \cite{wang2011enhanced}. A new SERS substrate, self-assembling 4-mercaptophenyl boronic acid (4-MPBA), on silica-coated graphene oxide with silver nanoparticles, has been published by Pham et al. \cite{zhu2012design}. The detection range of this substrate is 36–360 mg/dL1. These studies demonstrate CGM's potential.

\subsection{Photoacoustic}

Photoacoustic spectroscopy is another method of measuring glucose concentration in the blood without requiring a needle stick method. As a result, this apparatus uses infrequent bursts of laser light that are taken up by a particular molecule in the substrate to heat up a small area.  A quantum cascade laser (QCL) generates light that strikes the sample, resulting in an ultrasonic wave. This wave is on the move. The signal is sent from the acoustic resonator to the detector, where it is amplified, digitalized, displayed, and supplied to the computer \cite{unmussig2018non}.

Investigated a modified arrangement with two QCL sources, as depicted, to increase the sensitivity for in vivo glucose detection by covering one wavelength with considerable glucose absorption and the other with background wavelengths \cite{jiang2010highly}. Photoacoustic spectroscopy's potential in non-invasive CGM systems has been proven, but little testing has been done employing a unique CGM system. Clinical validation involves numerous investigations.

\subsection{Surface plasmon Resonance}

When light strikes a metal layer at the interface of media with different refractive indices, a resonant oscillation of conduction electrons occurs, as described by SPR \cite{unmussig2018non}. The laser light source transmits The electromagnetic field and polarized monochromatic light passing through a prism may be created by plasma free-electron-wave oscillations at specific resonance angles \cite{unmussig2018non}. Since the fluctuating electric field responds to variations in the medium's index of refraction, the peak of SPR resonance may be identified with variable frequencies of resonance derived from glucose concentration \cite{unmussig2018non}.

Constructed a reflective SPR optical fiber with GOx immobilized as the glucose-recognizing film \cite{kong2012graphene}. The enzymatic interaction between GOx and glucose caused the refractive index to change, which caused the SPR spectrum to shift with different glucose concentrations. With high selectivity and stability and a sensitivity of 0.85 nm mg1 dl1, CGM has a lot of potential in the future \cite{kong2012graphene}.

Presented a reflecting Integrating a glucose-sensitive SPR optical fiber membrane with SiO2-modified PAM gel This novel application of fiber optics with exceptional sensitivities, selectivity, and reaction speed, has produced intriguing results \cite{steiner2011optical,yetisen2014holographic} examined SPR optical fiber modified with boronic acid enhancement along with Au nanoparticles, which had a reduced detection limit and covered normal blood glucose levels. Table 7 compares SPR optical fibers' key features.

While SPR technology has found widespread use in chemical analysis in recent years, very little study has focused on its potential use for non-invasive glucose monitoring because of its great motion-sensing sensibility, perspiration, and temperature but low sensitivity to tiny glucose concentrations \cite{unmussig2018non}. Recent efforts, however, have been made to develop nanotechnology and materials science-based methods for making very optical sensors that are sensitive and refractory, with the hope of one day finding a solution to the problem.

\subsection{Holography}

Holography is a method of capturing three-dimensional images through the use of nanofabrication or other light-sensitive materials processes, and holographic gratings may be created by exposing a material to laser light \cite{vashist2012non}. Holographic sensors are a new area of optical sensing that is growing quickly. They may be able to finely bend light between the ultraviolet (UV) and near-infrared (NIR) ranges \cite{gifford2013continuous}.

When exposed to an analyte, holographic sensors' optical properties may change. Reflection holographic sensors and transmission holographic sensors are also examples of this \cite{vashist2012non}. Sensors based on reflection holograms may be employed transparently in the visible spectrum, with the changing wavelength of the diffracted light being readily noticed. Results for transmission holograms might be achieved using an optimal diffraction wavelength for a spectrometer that can be adjusted \textbf{}. when illuminated by a white light source, holograms can be used to calculate nanometer-scale distances between individual particles in mirrors.

where "n0" is the coefficient of refraction efficiency with respect to the medium for recording, "Lambda" is the distance between individual sections, and "theta" is the Bragg's angle as calculated by geometry \cite{vashist2012non,locke2018layer}. A lot of work over the last several decades has gone into developing holographic glucose sensing. It was shown that hydrogels based on phenylboronic acid might be used to create holographic glucose sensors that reversibly react to different glucose concentrations throughout the visible spectrum \cite{locke2018layer}. It was found that a hydrogel with a 20\% mole fraction of 3-actrylamidophenylboronic acid (3-APB) was the most selective and didn't interfere with lactate \cite{locke2018layer}.

In recent years, researchers have explored the use of 3-APB-established hydrogel to create a new holographic glucose sensor layout using double polymerization \cite{villena2019progress}. However, further work is needed to determine the appropriate response of the sensor. A delicately crosslinked phase (P1) is formed from a 3-APB monomer solution, followed by the formation of a substantially crosslinked phase (P2) from a second monomer solution \cite{villena2019progress}.

As was previously indicated, the pH interference may be reduced by being selective when applying indole-3-carboxylic acid derivatives. In light of this, researchers have looked at several boronic acid derivatives and found that 2-acrylamidophenylboronate shows promise as a contender with improved selectivity in the physiological pH range \cite{vashist2012non}. The colorimetric findings of in vitro investigations performed to quantify blood glucose levels within the visible range. Holographic glucose sensors have been shown to work, and they don't have any major delays or problems \cite{vashist2012non}.

Recent studies have examined the efficacy of holographic glucose sensors in both the visible and NIR ranges, with the sensors having been fabricated by hydrogel functionalized with 4-formylphenylboronic acid and laser ablation \cite{pickup2005fluorescence}. Utilizing a laser to vaporize a chitosan hydrogel enhanced with gold nanoparticles, a new holographic glucose sensor was developed that can detect the complete range of physiological glucose concentrations \cite{pickup2005fluorescence}. ISFs' fructose, vitamin C, and lactate were shown to have little effect on the sensor's performance, and the sensor was determined to be insensitive to changes in a measure of temperatures, pH, or ionic strength \cite{pickup2005fluorescence}. Hydrogels modified with boronic acid were used. thought to be extremely sensitive and biologically compatible, showing encouraging findings for the prospective use of implantation, but more in vivo validation is essential.

Optical coherence tomography, polarimetry, thermal infrared spectroscopy, photoacoustic spectroscopy, and Raman spectroscopy are examples of noninvasive optical methods without the use of fluorophores \cite{Krans2018,wang2015recent}. Skin pigmentation is not a factor in the optical detection of glucose using near-infrared (NIR) spectroscopy, which allows 90–95\% of the light to pass through the human stratum corneum and epidermis into the subcutaneous region with little tissue adsorption. Changes in glucose concentrations cause changes in the subcutaneous tissue's dielectric strength, polarizability, and permittivity, which in turn affect the NIR absorption, reflection, and refraction in the near infrared \cite{Krans2018,shibata2010injectable,heo2011long}. The greater absorptivity of glucose has a detrimental effect on the measurement's sensitivity and selectivity because water has a prominent absorption band in the NIR and light scatters in the tissue \cite{pickup1999vivo}.

Similar to this, by measuring inelastically scattered photons, glucose concentrations have been determined non-invasively using Raman spectroscopy \cite{tura2007non,cunningham2009vivo}.The vibrational or rotational energy of chemical bonds in the system (such as those linked to glucose, for example) is inversely correlated with the energy changes from these photons Temporary delays in spectrum capture and stabilization are required, even though Raman spectroscopy produces narrow and distinct peaks (in contrast to NIR absorbance) \cite{MayoClinic2017}.Recent research has shown that surface-enhanced Raman spectroscopy (SERS) significantly reduces acquisition times while increasing sensitivity and detection limit \cite{barman2010accurate,lyandres2008progress}.Van Duyne and colleagues have shown that SERS is a more selective method of measuring glucose on an AgFON was coated with a (1-mercaptoundeca-11-yl) tri(ethylene glycol) (EG3) partitioning layer to preconcentrate glucose within an electromagnetic field enhancement zone \cite{lyandres2008progress}.

In order to use a SERS biosensor in a clinical setting, a device that exposes nearby tissue to continuous laser radiation risks having photothermal damage \cite{MayoClinic2017}.Other difficulties with using nonfluorophore-based optical exposure include response effects brought on by motion, tissue heterogeneity, pH, and temperature, which have a negative impact on the accuracy and selectivity of blood glucose measurements \cite{wang2015recent,tura2007non,shafer2003toward}.

Other techniques for measuring glucose besides the conventional electrochemical or optical methods include impedance and electromagnetic spectroscopy. Impedance spectroscopy, which transports alternating current through the tissue, can be used to identify changes in plasma conductivity as a function of glucose concentration \cite{alexeeva2010impact}. The plasma's sodium and potassium concentrations change as a result of a rise in the local glucose concentration, which also affects the erythrocyte cell membrane's dielectric strength, permittivity, and conductivity \cite{alexeeva2010impact}.Unfortunately, poor accuracy is caused by the fact that glucose is not the only factor affecting changes in blood dielectric properties. Furthermore, it has been demonstrated that such measurements are adversely affected by the body's temperature and state of illness \cite{alexeeva2010impact}.

\section{CGM as a replacement for SMBG}

Professionals were first given access to standalone CGM systems, but today consumers may purchase devices that perform the same functions as SMBG and so achieve personal real-time CGM. The Food and Drug Administration approved the Dexcom G5 Mobile for sale in 2016. A Continuous Glucose Monitoring System to Replace Fingerstick Decisions on diabetes therapy are based on blood glucose testing for people with diabetes aged 2 years and older. This was the initial CGM system to be authorized by the FDA for use in managing diabetes without first confirming results with a fingerstick test. In making treatment choices for diabetes, the method was allowed to work in conjunction with, rather than as a substitute for, traditional finger-stick testing. Then, the Abbott Freestyle Libre continuous glucose monitor (CGM) was authorized as the sensor to substitute blood glucose monitoring through fingerstick in making therapeutic choices for people with diabetes.

9: Sensor and system architectures for continuous glucose monitoring (CGM), wearable glucose sensors (GSS), and self-monitoring blood glucose (SMBG), including sample sites and device components.
Figure

\section{Glucose monitoring in real time}

CGMs, which measure blood sugar levels in real time, have been on the market for a while; most use an electrochemical method including a glucose-sensitive enzyme, while several microdialysis-based CGM systems are also on the market.

Currently available electrochemical CGM systems, like those from Abbott, DexCom, and Medtronic, allow for the subcutaneous implantation of an enzyme-promoted electrode, which then transmits data wirelessly to a digital receiver in real time \cite{lee2021continuous}. Rather than obtaining a sample of blood from a capillary, CGM devices typically monitor the glucose level in the ISF with a less intrusive outcome \cite{sun2021improving}. This is because it has been shown that equivalent findings may be achieved when the glucose level is stable. While this method would allow for continuous monitoring, it would still need periodic finger-prick tests to properly calibrate the CGM and account for fluctuations in glucose levels \cite{sun2021improving}. In addition, most of the CGM devices that have been authorized up to this point are still disposable and will likely need to be changed every week \cite{wild2004global}.

There has been a lot of focus on sensor development as a way to greatly increase the lifetime of sensors in recent years because of the widespread pursuit of better electrochemical biosensors. For example, the Eversense sensor developed by Senseonics has been successfully applied to a completely implanted CGM device, which can deliver real-time data for glucose levels for a span of 90 days. This device has been authorized for the European market and prompted new research concentrating on optical glucose monitoring, while sensor biocompatibility and patient acceptability remain important challenges for implanted systems \cite{ehrhardt2011effect}.

An overview of currently available commercial CGM systems is shown. The accuracy of the following systems is measured by the mean absolute relative difference (MARD), which is the average of the absolute inaccuracy among all values acquired from the CGM system and matching corresponding values \cite{bard2022electrochemical}. As a result, it could be described as a percentage, with a lower MARD value indicating higher device accuracy \cite{bard2022electrochemical}.

\section{Glucose-sensing methods}
\subsection{Microwave}

The ability of microwaves to achieve the point where a more precise glucose measurement may be achieved simply and quickly with appropriate blood concentration has garnered a lot of interest \cite{unmussig2018non}. Reflections, transmissions, and resonant perturbation are three fundamental microwave concepts that might be used in glucose monitoring using a near-field sensor \cite{shibata2010injectable}. By measuring the entire set of S parameters, including the transmission parameter S21, transmission methods enable sophisticated calculations for more precise findings \cite{unmussig2018non}.	The reflection method measures the reflection parameter, S11, to determine the amplitude and phase variation caused by the changing permittivity of blood with various glucose concentrations. On the other hand, the resonant perturbation technique \cite{unmussig2018non} accounts for correlation change through measurements of a 3 dB bandwidth, a quality factor, and resonant frequency. All these techniques, however, are restricted to a single operating frequency and are very sensitive, making them susceptible to interference from environmental shifts.

Used a Hilbert-shaped microwave sensor to detect glucose concentration in water-based solutions at a frequency of 6 GHz with a minimum detectable amount of 1.92 mg (Odabashyan et al.) \cite{aloraefy2014vitro}.In addition, the sensor's temperature dependency has been evaluated, and respectable temperature correlation values have been achieved for concentrations between 50 and 150 mg/dL.

If researchers want to make progress toward their goal of non-invasive CGM (constant glucose monitoring), it is crucial that their glucose sensor be described not just in dissolved solids but also in simulated biofluids like sweat or ISF. Investigated the functionality of a chipless tag split-ring resonant with simulated ISF approaches, achieving a detection scale of 36–450 mg dl1 and a detection threshold of 0.01 mg dl1 at an operating frequency of 4 GHz \cite{klonoff2012overview}.

However, a nanostrip-based microwave biosensor has been described by for non-invasive glucose observation through sweat \cite{freeman2009competitive}. As can be seen, the sensor remained improved over sensors that detect microwaves using microstrip antennas by having GOx immobilized on the nanostrip. Key differences and similarities between these sensors are compared.

\subsection{Radiofrequency}

Detection of high-sensitivity glucose in the absence of a facilitator has garnered a lot of interest in radiofrequency (RF)-based biosensors \cite{smith2015pursuit}. A glucose sample's dielectric characteristics may be acquired and associated with the sample's glucose concentration in a manner similar to microwave resonators by determining the fluctuation in resonance frequency with modifications in signal amplitude \cite{li2015u}. Non-invasive glucose monitoring with this technique is seen as promising since it might solve the problems of environmental interference and short lifespan that plague conventional electrochemical glucose sensors \cite{mohammadi2014vivo}. It is possible to produce a rapid response using RF biosensors for label-free glucose measurement, and no pre-stabilization is necessary \cite{mohammadi2014vivo}. However, the absorption of liquids poses a big problem for this technique since radio signals would be disrupted. In addition, prolonged exposure to RF waves poses a serious threat to human health. As a result, additional research is required to lessen the danger \cite{qi2016glucose}. Some of the most recent accomplishments using this technique.

\subsection{Bioimpedance}

Bioimpedance, which quantifies electrical impedance passing in the body's connective tissues of a live creature, has been thought about as a feasible option for non-invasive glucose monitoring \cite{sharma2016bisboronic}. Changes in blood glucose level would cause a shift in the ratio of potassium to sodium ions inside red blood cells, which in turn would alter the membrane's permittivity and conductivity \cite{unmussig2018non,sharma2016bisboronic}. Because of this, bioimpedance spectroscopy was developed to detect the associated resistance using an alternating current, from which the conductivity might be derived and promptly connected by using glucose content \cite{unmussig2018non}. Although this technique requires less time and money to implement, its extreme sensitivity to heat and perspiration is cause for concern

\cite{sharma2016bisboronic}. Additionally, the purely physical circumstances of the cellular membranes may have a major impact \cite{unmussig2018non}. Although it was shown to be able to linearly connect glucose concentration and electrical characteristics with just 3.7\% accuracy, further work is needed to see it used in non-invasive CGM \cite{sharma2016bisboronic}.

Figure 12: Method of measuring tear glucose levels using a contact lens biosensor

\subsection{Sensing Design and New Materials}

Quantum dots (QDs) made of semiconductors have been widely researched as photoluminescent nanomaterials for use in sensing and cell imaging due to their high performance and adjustable size-depending optical characteristics \cite{badugu2005glucose}. By combining photoelectrochemical QD sensors with the right enzymes, glucose may be detected indirectly in samples \cite{shibata2010injectable,heo2011long}. The glucose signaling chain from glucose oxidase to molecular oxygen to CdSe/ZnS QDs at the electrode that used to demonstrate indirect sensitive detection of glucose Tanne et al. \cite{pickup1999vivo}.

The photocurrent was used to measure glucose oxidase's enzymatic activity, which catalyzes the process of glucose oxidation through the absorption of O2. O2 consumption during lighting resulted in a lower photocurrent. Adjusting the quantity of enzyme atop the QD layer was discovered to have a significant impact on the sensing capabilities of this sort of electrode, and this could be done using the layer-by-layer method. Glucose, glucose oxidase, oxygen, and quantum dots (QDs) form the signal chains upon which the aforementioned glucose sensors operate. A shuttle molecule may tie the enzyme to an electrode, allowing for facilitated electron transfer from the biocatalyst. By alternating the deposition of water-soluble QDs made of CdSe-CdSe with a combination using a TiO2 electrode containing [Co(phen)3]2+/3+ and poly(ethyleneimine), Zheng et al. The decreased enzyme transferred charge carriers to the electrode, and the resulting photocurrent was related to the glucose concentration.

Current promising solutions for glucose monitoring include label-free optical detection systems that can be used in real time. Imagine CMOS transistor (complementary metal oxide semiconductor) sensors being transformed into effective instruments for biosensing applications due to their excellent high signal-to-noise ratio, minimal energy use, downsizing, ease of smartphone access, and smoother integration into tiny diagnostic tools in medicine. As an added bonus, CMOS image sensors may use a wide variety of glucose monitoring techniques, such as continuous glucose monitoring enzymatic glucose monitoring and thermal glucose monitoring \cite{cunningham2009vivo,lyandres2008progress,shafer2003toward}.

An image sensor using complementary metal oxide semiconductor technology for very sensitive glucose detection through variations in basic photon count. Enzymatic oxidation of glucose was catalyzed by adding glucose in quantities ranging from 100 to 1,000 milligrams per deciliter onto a basic PDMS (polydimethylsiloxane) microchip Kim et al. \cite{barman2010accurate}. The enzymatic interaction of oxidized glucose with a chromogen results in a brown hue, the intensity of which varies with the quantity of glucose. The sensor receives photons that have passed through a PDMS chip and have a range of spectral densities. The CMOS image sensor detected the number of photons based on color saturation in relation to the amount of glucose in the blood. This information in relation to the amount of glucose in the blood in relation to the amount of glucose in the blood in relation to the amount of glucose in the blood in relation to the amount of

glucose in the blood in relation to the amount of glucose in the blood in relation to the amount of glucose in the blood in relation to the amount of glucose in the blood in relation to the amount of glucose in the blood in relation to the amount of glucose in the blood in relation to the amount of glucose in the blood. It is feasible to evaluate a broad scale of BG levels with remarkable linearity based on the CMOS image sensor, and this approach might thus promote a simple point-of-care diagnostic method by matching the acquired digital data with glucose concentration.

\section{Development of New Biocompatible Sensor Coatings}

Adsorption of proteins at surfaces and interfaces is the main obstacle in designing materials that come in contact with tissue for bioaffinity and medical implant sensors. Coatings that provide protection may significantly interact with proteins surrounding most minimally invasive biosensors, reducing the likelihood of tissue responses caused by device implantation \cite{yonzon2004glucose}. The antifouling material's structure should be designed to be hydrophilic while remaining electrically neutral and to accept hydrogen bonds rather than supporters in order to produce a surface that is impermeable to proteins. Antifouling formulations for sensing applications with new architectures and non-specific protein resistance are designed in accordance with these needs \cite{yonzon2004glucose}. Organic/inorganic composites and biofunctional polymer architectures are only two examples of the many promising options that have been found. Enzymes have been shown to work well with silicates generated from sol-gel processes \cite{alexeeva2010impact}. Coatings for glucose oxidase-based sensors made from silica-based hybrid materials have been found to be non-toxic and biocompatible, with good in vitro and in vivo biocompatibility. In both BSA buffered solutions and serum, the silica-polymer coated sensors demonstrakted stable glucose responses \cite{caduff2006non,vashist2012non}. Unfortunately, no mention was made of the effectiveness of these devices in preventing fouling.

In the present day, implanted CGM sensor coatings are made using a wide variety of macromolecules, some of which are natural, some of which are semi-synthetic, and some of which are synthetic (cf. Table 1). Natural materials like alginate, derivatives of chitin, chitosan, and related polysaccharides, are extensively studied because they are equivalent to natural precursors and the biological environment is equipped to detect and deal with their metabolites for biosensors in the form of layers, membranes, and coatings \cite{unmussig2018non}. In contrast, customized fabricated polymers with specialized architectures and finishes frequently perform better than conventional macromolecules, particularly in terms of decreased immunogenicity. Poly (lactic co-glycolic acid), poly(ethylene glycol), poly(lactic co-glycolic acid) (hydroxyethyl methacrylate), poly(vinyl alcohol), and others are biocompatible and biodegradable polymers \cite{liu2015high}.Polymers generated from HEMA and PEG are the first and second generations of effective protein-resistant compounds, respectively \cite{liu2015high}. They now have a variety of applications and may be used to coat mini-invasive devices to alter their behavior. When used alone or in conjunction with other polymers, they are also widely used as hydrogel matrices \cite{liu2015high}.

Hydrogels are semi-open structures made up of entangled chains that are hydrophilic, water-insoluble, and very absorbent and storable in water. They are also highly permeable to tiny molecules. Hydrogel coatings are used because glucose may diffuse through the swelling hydrogel layer. Physical and chemical manipulation of the hydrogel's cross-linking density, which determines the hydrogel's water content and, by extension, the polymer network's mechanical strength, allows for fine-tuning of the glucose diffusion rate. The outer layers of a biocompatible polymer hydrogel have been shown to improve the efficacy of a mini-invasive needle-type glucose sensor in both laboratory settings and animal studies.

PEG-based antifouling matrices are well-known and widely accepted and have been extensively studied \cite{wei2020transfer}, despite oxidation issues that may compromise the antifouling characteristics with prolonged use. Poly (ethylene glycol), methyl ether methacrylate (PEGMA), and other similar substances were used to create hydrogels. Studied in the same way as polyethylene glycol (PEG) for the development of modern anti-biofouling materials \cite{wei2020transfer}. Because of this, the results demonstrated that the total hydrophilicity of the gel was directly controlled by the molecular weight of the PEGMA homopolymer, which was found to have an intermediate hydrophilic chain length and an intermediate connecting degree. Furthermore, microbiofouling caused by bacteria and blood cells revealed that PEGMA hydrogels provided the best non-fouling capability, implying that they could be used in the coating of blood-contact devices and the regulated release of drugs from non-fouling hydrogels in living organisms.

\subsection{Salivary Biosensors}

Several studies reveal that diabetics have more glucose in their saliva. These studies aimed to explore whether non-invasive salivary glucose monitoring might help control diabetes. Saliva is a good diagnostic fluid because blood components may be absorbed via the mucosa \cite{mohammadi2014vivo}. In both healthy and diabetic patients, saliva glucose concentrations range from 20 to 200 mmol/L \cite{mcnichols2000optical}.Although levels of glucose and carbohydrates in the saliva are somewhat correlated as a group \cite{qi2016glucose}\cite{sharma2016bisboronic}, the correlation is much stronger, allowing for BG levels to be monitored in real-time across many sites within the same predicted glucose levels in the saliva \cite{pham2017glucose}.

Scientists are working on saliva-based glucose monitoring devices. Quantified chemicals using colorimetric testing Lambert et al. \cite{kottmann2016mid}. Some studies have utilized Clarke-type electrodes to monitor salivary glucose \cite{mcnichols2000optical}, while others, like Lipson, have used NIR or Raman spectroscopy \cite{zheng2020highly}. Immobilized strips of filter paper containing glucose oxidase enzyme (1.4 U/strip) and reacted them in the presence of a color pH indicator with synthetic glucose samples (Soni et al.) \cite{sharma2016bisboronic}. The presence of glucose in the reaction medium caused a color change in the filter paper, which could be scanned (using RGB profiling) and studied using Free and Open-Source Image Editing Software (GIMP).

\section{US-based CGM products are currently on the market}

\subsection{Dexcom}

The Dexcom glucose sensor uses an electrochemical approach that's outdated. The sensor uses a platinum or platinum iridium "operational" electrode and a "reference/counter" electrode made of Ag/AgCl. A functional electrode oxidizes hydrogen peroxide. To create the sensor's circuitry, a platinum or platinum iridium wire is patterned with insulating polymers via photolithography, grit blasting, etching, and laser ablation (Simpson et al., 2015). The operating and reference/counter electrodes can be insulated with polyimide, polyurethane, or perylene (Simpson et al., 2015). This allows regulated exposure of the working electrode beneath. Tapsak et al. (2002) say that the Ag/AgCl layer for the electrode used as a point of comparison is put on by dip coating, screen printing, or jet printing in order to get the right thickness and location.

After the electrode layers are produced, the Dexcom sensor chemical is applied (via photolithography, dip coating, etc.). Additional chemical layers may be divided into "domains," each containing a unique set of sensors. In electrochemistry, active enzymes are contained inside the enzyme domain. GOX is a Dexcom enzyme (Broock and Rixman, 2009). In addition, the layer will comprise a polymer that may cross-link, such as perfluorocarbons or silicon covered with polyethylene glycol, to inhibit enzyme escape and assure ample oxygen for enzyme reactions. Next, a glucose diffusion barrier layer is added to ensure the sensor operates in conditions where limiting diffusion is present and the enzyme layer has adequate oxygen. Broock and Rixman (2009) say that the diffusion membrane is made of a polymer that lets some oxygen through but not glucose.

\subsection{Medtronic}

Medtronic's insulin monitors are wireless and use first-generation glucose measurements. Medtronic's sensor contains three electrodes: a functioning, a counter, and a reference electrode. Platinum is used for both the working and counter electrodes and electroplated to the desired roughness. The surface reference electrode is Ag/AgCl (A number of authors have discussed this topic, including Shah and Gottlieb (2005), Van Antwerp and Mastrototaro (1999), and Cheney and Van Antwerp (1994). Unlike Dexcom, the counter and reference electrodes are distinct. Medtronic's reference electrode is used as a voltage regulator and current meter, with the counter electrode providing the reducing molecules necessary to maintain a steady current, like oxygen. A carrier substrate covered with polyimide is employed as the electrode foundation in sensor design. Photolithography, both moist and dry etching, and evidence are then used to pattern the three electrodes layer by layer (Van Antwerp, Mastrototaro, 1999).

After substrate layers, Medtronic applies synthetic chemistry for a glucose sensor. These domains contain GOx and glucose diffusion. The enzyme domain is placed by dipping, spraying, etc. The addition of glutaraldehyde then creates a domain of adhesion-like crosslinking to avoid enzyme dilution and decrease substrate delamination. The domain of glucose limitation employs polyamines and siloxanes to generate a glucose-semipermeable barrier. Siloxane is not glucose permeable, whereas polyoxypropylene-diamine is (Shah and Gottlieb, 2005). Changing these polymer ratios controls glucose transport via the limiting region. The glucose-limiting domain prevents enzyme saturation and ensures enough oxygen (Van Antwerp 1995).

\subsection{Abbott}
Abbott's glucose sensor works like Medtronic's. Abbott employs second-generation technology. The complex of osmium acts as an intermediary for electrons on the surface of GOx and carbon electrodes. Then, Abbott's sensor architecture is similar to Medtronic's, having three separate electrodes—functioning, counter, and reference. The Ag/AgCl electrode reference provides a stable overpotential.

Using the electron mediator, the osmium complex makes an oxygen-free and self-contained sensor of tissue oxygen. Osmium complexes have a lower oxidation potential than H20 peroxides. This allows Abbott to apply a reduced overpotential in relation to the working electrode, minimizing some interference (Say et al., 1998; Heller and Feldman, 2008). (Heller and Feldman, 2008; Say et al., 1998). Abbott's sensor is manufactured with

\subsection{Senseonics}

Unlike its competitors, Senseonics does not rely on electrochemical sensing but rather on fluorescence detection based on the interaction of a divalent boronic acid coupled to a fluorophore with glucose. Photoinduced electron transfer (PET) from the unbound boronic acid reduces fluorescence at low glucose concentrations, whereas interactions between glucose molecules boost the fluorescent signal at high concentrations. Boron-nitrogen bonds are formed, preventing PET from developing in the amine, and this is the mechanism at work here.

In contrast to enzyme-based sensors, a product is not formed when glucose and boronic acid join; only a bond that is proportional to the glucose concentration is formed (Mortellaro and DeHennis, 2014). The molecular recognition element layer is coated with poly (methyl methacrylate) (PMMA) to regulate glucose flow and ensure biocompatibility.A silicone ring containing dexamethasone (DEX), which is released at controlled rates, reduces inflammation in the area around the sensor. This makes the sensor last longer.


\section{CGMs and associated technology currently on the market}

The development of glucose sensors and associated technologies has continued along with the rise in the number of diabetes patients. As a result, new insulin analogs and a variety of continuous glucose monitors have been created and are now commercially accessible to enhance treatment and results. Patients with either type I or type II diabetes have greatly benefited from extensive research and commercial success with continuous glucose monitoring \cite{carlson2017clinical,gorst2015long}. When type I diabetes is diagnosed early on and requires extensive insulin treatment, the CGM has emerged as the gold standard \cite{laffel2020effect,beck2017effect}. Improvements in HbA1C results have been seen by the Juvenile Diabetes Research Foundation Continuous Glucose Monitoring Study Group \cite{nierras2010juvenile}. By contrasting the usage of a CGM vs. a SMBG (>4 tests per day) during a 24-week period, the DIAMOND research demonstrated a greater improvement in A1C of 1\% as opposed to 0.4\% \cite{beck2017effect}. As a result of the comparable DIAMOND trial, in which 158 individuals were diagnosed with type II diabetes, CGM usage has grown in both type II and prediabetic patients. This is because CGM use resulted in an average HbA1c improvement of roughly 0.3\% in comparison to the nominal treatment \cite{beck2017effect}. In a short study with 32 prediabetic patients, Chakarova et al. (2019) found that people with prediabetes had more glucose variability than non-pre-diabetics, indicating that it may be suitable to use CGMs for glucose monitoring sooner than previously thought. The CGM's many application cases have advanced the technology.

These breakthroughs paved the way for CGM usage in a wide range of populations. Following a description of the four commercially accessible CGM systems and their sensor efficiency specifications, this part introduces the sensor structures of the CGM systems that are already commercially available, depending on each different sensor domain. Then, the algorithms, calibration, filtering, and current causes of inaccuracy in sensors that measure glucose levels in real time are described.

\section{Future Prospective}

Due to its simplicity of operation and greater comfort in comparison to SMBG-type sensors, future projected CGM systems are likely to see an increase in utilization among people with Type I and Type II diabetes. Despite improvements in CGM structure accuracy, BGM sensors continue to provide more accurate results than CGM systems \cite{campos2017effect,heinemann2018real,freckmann2018measurement}. Additionally, the present cost factors of CGM need to be addressed and enhanced if type II diabetes patients are to utilize CGM devices more often. A CGM system would function best if it didn't need implantation. Two decades ago, Cignas Inc. and Animas Inc. developed and marketed Glucowatch, a cutting-edge continuous glucose monitor that uses an enzyme-based sensor and a non-contact sampling mechanism. The technique used reverse iontophoresis to measure the glucose that was taken out of the ISF. An adhesive that stayed on the electrode's skin for three days allowed for ongoing monitoring. However, the system was abandoned in 2007 as a result of customer complaints regarding its accuracy and the way it sometimes caused skin discomfort. Recently, Nemaura Medical introduced the SugarBEAT CGM system, which was marketed in the UK.

It used a Glucose sampling that is non-invasive technology comparable to Glucowatch, although with more features and a shorter use period. It also replaced the sensor patch daily, preventing skin irritation. Despite the fact that perspiration, tears, urination and saliva are excellent materials for Glucose sampling that is non-invasive monitoring, their clinical use is still not widely acknowledged. The interval between each sample and the blood glucose level is one of the causes. According to Heikenfeld et al. \cite{heikenfeld2019accessing}, saliva, perspiration, and tears all exhibit time lags of one to ten minutes. The tear glucose time lag has also been according to be five to ten minutes \cite{badugu2005enhanced}. Additionally, the concentration of perspiration, saliva, and tears is over a thousand times lower than that of blood glucose levels, and it is even lower than that of ISF, which has a concentration that is up to ten times lower than that of blood glucose levels \cite{heikenfeld2019accessing}. Using those sample sites in a clinical setting is difficult due to the murky relationships with blood glucose. Although more thorough research is needed before a definite association between blood glucose and sweat, tears, and saliva can be established, we think that those samples may be employed for continuous glucose monitoring in the future and will proceed in the same manner as ISF. On the other hand, urine not only lags behind blood sugar levels but it also contains typical renal a threshold of approximately 180 mg/dl. However, diabetes patients' renal maximum threshold values range from 54 to 300 mg/dl, and as a result, the fluctuation in the renal maximum threshold results in a misreading of the real glucose level in the blood \cite{walker1990clinical}. In other words, if the renal threshold is changed, a diabetes person with a glucose level in the blood of 150 mg/dl may get a reading of 54 mg/dl. Due to the problems that urine poses, it is thought to be challenging to utilize urine as a trustworthy CGM sample. Urine glucose content is a poor indicator of glycemic level owing to the widespread use of SGLT2 inhibitors, or sodium-glucose cotransporter-2 inhibitors, reduce blood sugar levels by forcing sugar removal by the kidneys through urine. The resurgence for glucose detection via reversed iontophoresis proposes innovative and efficient ISF sampling that is non-invasive techniques in the near future, given that glucose concentrations method non-invasively accessible samples are currently not clinically relevant.

Due to the popularity and effect of the Eversense Continuous Glucose Monitoring System (Senseonics), researchers and businesses have turned their attention to completely implanted chronic CGM systems that can operate continuously in vivo for months or even years. Non-allergic adhesive materials are now available, rather than the molecules of the sensor or enzyme, which limit the operating time of the sensors that are implanted beneath the skin and controlled along with the device put on the skin. which, over the course of the attachment's two weeks, might create skin irritation as well as sanitary problems \cite{englert2014precision,messer2018preserving,gisin2018manufacturing}. Instead, the Eversense Continuous Glucose Monitoring Method used controller components that were entirely implanted with the sensor. Since the sensor is driven by near-field communication (NFC) technology, implantation of the sensor should be carried out by a skilled surgeon using microsurgery. Patients should see their doctor every three months to start and change the sensor or gadget. Due to COVID-19, which made it difficult to contact doctors, this necessity for very frequent doctor visits led to a temporary halt of Eversense CGM commercial sales in the United States. While recognizing cutting-edge microelectronics and low-power wireless communication systems, research on completely implanted electrochemical sensors has also been described. Gough, who invented the Glysens glucose sensor, is an example of this. It assessed the loss of O2 just like the original Clarke glucose sensor. A secondary enzyme called catalase was used in the Glysens system's in vivo testing on dogs for more than a year in order to lessen the effects of GOX inactivation brought on by H20 peroxide \cite{armour1990application,gough1982stress}. The Eclipse 3 ICGM Device, a completely implanted system, is now the subject of a two-year clinical research study being conducted by Glysens \cite{Glysens2023}.

So, to make continuous long-term chronic CGM systems possible, a new technological advance is expected not only in sensor principle and sensing molecule development, but also in methods for sensor implantation and removal without professional surgical processes, as well as in the development of non-allergic or minimally allergic and antibacterial materials for long-term sensors and controlled adhesion on the skin.

Researchers are working on the next generation of biomedical devices for Type 1 diabetes patients, this is a completely functional artificial pancreas, as a consequence of the accomplishment of realizing insulin pumps with built-in sensors, which remain controlled based on the findings of the CGM system. Closed-loop technologies now in use monitor blood sugar levels and provide the necessary dose of insulin automatically. Recently, a technique was developed to simultaneously inject the hormones glucagon and insulin. These techniques employ complex algorithms to determine the ideal insulin dosage to inject and then tell an insulin pump how to administer it. In order to closely regulate the patient's glycemic response, current difficulties include progress toward non-glucose information sensing, particularly physiological factors.

Blood glucose levels must be closely monitored, which is difficult due to the intricacy of glucose homeostasis.

The metabolic and physical activities of a patient have a big impact on the dynamics of glucose and insulin.

As a result, tracking energy use may serve as a proxy for physical activity. Physical exercise may raise the risk of hypoglycemia because physiological stress affects a patient's blood sugar level. Therefore, to manage closed-loop investigations, physiologic stress must be taken into account. Galvanic skin response (GSR) and stress have a known association; hence, GSR may be employed as a stress indicator in conjunction with closed-loop system control. Additionally, ongoing insulin concentration monitoring will enhance the tight control of blood sugar levels and reduce the danger of hypoglycemia. As a result, substantial research is now being done to integrate CGM systems with other wearable monitoring systems for physiological parameters \cite{turksoy2018multimodule}. CGM systems, for example, will be heavily reliant on the combination and availability of sensors for both passively observing physical activities and developing a completely automated artificial pancreas system, necessitating the use of less invasive, constantly monitoring physiological indicator biosensors.

In terms of green innovation that reduces carbon footprint, the development of CGM systems will not be an outlier.The nonprofit group Diabetes Technology Society (DTS) is dedicated to advancing the purpose of technology in the battle against diabetes. It has pioneered the introduction of CGM systems and is leading the charge to hasten the development of artificial pancreas devices. The Green Diabetes Initiative, recently launched by DTS \cite{klonoff2020diabetes}, aims to reduce waste and environmental resource consumption while promoting environmental sustainability. Used disposable diabetes devices meant for single use in the house generate a lot of garbage. Injection needles, syringes, lancets, continuous glucose sensors, blood glucose monitors Insulin containers, scripts, infusion tubes, consumable pumps, device batteries, and packaging are some examples of these gadgets. They demand that we reduce, reuse, recycle, redesign, and re-educate ourselves about diabetic device waste. These five "R" methods are designed to help you get the most out of diabetes in a disposable way. As a result of this push, CGM system manufacturers have been working hard to reduce the size and packaging of their devices in order to comply with the Green Diabetes Initiative. Future improvements to CGM systems and their sensors should support this initiative by not only shrinking their size but also altering their design, materials, and operational stability, as well as making electronics, sensors, transducers, chemicals, and molecular biosensors more readily available for reuse and recycling.

\section{Conclusion}

Tight glycemic control is essential for the optimum management of diabetes mellitus because it lowers the risk of microvascular difficulties. The best monitoring methods for glycemic management in diabetic patients are now continuous glucose monitoring (CGM) devices. The first CGM system, the first glucose sensor for personal use that could be bought, and the first idea for a glucose sensor were all found in the US.

The development of glucose sensors suited for CGM applications has been claimed to have faced a variety of difficulties. The electrochemical principles used in the recent commercially available CGM systems (Medtronic, Dexcom, and Abbott) are the main ones for glucose sensing. These systems use the gold-standard enzyme glucose oxidase as the glucose sensing molecule in conjunction with hydrogen peroxide monitoring or with the addition of a hydrogel that harbors redox mediators. A CGM for long-term use was just put on the market. It uses a synthetic abiotic receptor with a fluorescence probe and a fluorescent detecting device (Senseonics).

The alerting systems' most notable feature is that they are part of the CGM system, which measures hyper-and hypoglycemic levels.This suggests that the current CGM system is capable of continuous in vivo and local glucose monitoring, as well as forecasting future glucose concentrations by analyzing glucose permeability over time.A hybrid closed-loop system, a medical device for insulin infusion in Type I diabetic patients, has been developed as a consequence of the improvement in glucose monitoring accuracy. Patients may now adjust their diabetes treatment plan based only on CGM data.

Glucose in tears, sweat, saliva, and urine is now a popular target for the development of less invasive or non-invasive monitoring devices. The field of biosensors is seeing a boom in interest right now, and wearable biosensors include these devices. The therapeutic importance of glucose in tears, perspiration, saliva, and urine has been documented by several wearable glucose sensors. These sensors transmit data using cutting-edge microsystems like wireless and self-powered networks and collect information using novel microsystems including micro-needle arrays, transducers, and flexible electrode-based sensing platforms.

The development of CGM systems may be broken down into four broad classes: 1) the creation of non-allergenic or slightly allergic materials, as well as antimicrobial materials; 2) the development of fully implantable chronic CGM systems based on a technique of sensor insertion and removal that does not need expert surgical procedures; and 3) the use of CGM systems as BG system replacements for Type 2 diabetic patients that is extremely accurate, less intrusive, and cost effective

\bibliographystyle{unsrt}  
\bibliography{references}

\end{document}